\documentclass[preprint]{aastex}

\newcommand{\nablaa}{\nabla_{\alpha}}
\newcommand{\nablab}{\nabla_{\beta}}
\newcommand{\pad}{\partial}

\newcommand{\beq}{\begin{equation}}
\newcommand{\eeq}{\end{equation}}
\newcommand{\beqn}{\begin{eqnarray}}
\newcommand{\eeqn}{\end{eqnarray}}
\newcommand{\lppr}{\stackrel{<}{\scriptstyle \sim}}
\newcommand{\gppr}{\stackrel{>}{\scriptstyle \sim}}

\slugcomment{ApJ, to appear}

\begin{document}


\title{Shear Acceleration in Expanding Flows}


\author{F.M. Rieger}
\affil{ZAH, Institut f\"ur Theoretische Astrophysik, Universit\"at Heidelberg, Philosophenweg 12, 69120 Heidelberg, and} 
\affil{Max-Planck-Institut f\"ur Kernphysik, P.O. Box 103980, 69029 Heidelberg, Germany}
\email{frank.rieger@mpi-hd.mpg.de}

\and

\author{P. Duffy}
\affil{University College Dublin, Belfield, Dublin 4, Ireland}
\email{peter.duffy@ucd.ie}

\begin{abstract}
Shear flows are naturally expected to occur in astrophysical environments and potential sites of 
continuous non-thermal Fermi-type particle acceleration. Here we investigate the efficiency of 
expanding relativistic outflows to facilitate the acceleration of energetic charged particles to 
higher energies. To this end, the gradual shear acceleration coefficient is derived based on an 
analytical treatment. The results are applied to the context of the relativistic jets of active galactic
nuclei. The inferred acceleration timescale is investigated for a variety of conical flow profiles (i.e., 
power law, Gaussian, Fermi-Dirac) and compared to the relevant radiative and non-radiative loss 
timescales. The results exemplify that relativistic shear flows are capable of boosting cosmic-rays 
to extreme energies. Efficient electron acceleration, on the other hand, requires weak magnetic 
fields and may thus be accompanied by a delayed onset of particle energization and affect the 
overall jet appearance (e.g., core, ridge line and limb-brightening).
\end{abstract}

\keywords{Outflow, jets: general -- Particle acceleration: shear -- AGN} 

\section{Introduction}
The non-thermal radiation seen from astrophysical objects bears witness to the presence of energetic 
charged particles that have experienced efficient acceleration within these sources. In the galactic 
domain, new high-resolution observations of supernova remnants have brought fresh momentum to 
the theory of diffusive shock acceleration \cite[e.g.,][for review]{Bell13}, while short-term variability 
seen in the context of active galactic nuclei (AGN) has motivated deeper studies of one-shot (gap- 
or reconnection-type) particle acceleration scenarios \cite[see e.g.,][for review of the case of M87]
{rieger12}. Complementary, new observational results in the radio and VHE domain and progress in 
our understanding of turbulence modeling have given new impetus to turbulent shear acceleration 
and emission scenarios \cite[e.g.,][]{aloy08,Sahayanathan09,liang13,grismayer13,ohira13,laing13}. 
The present study focuses on the potential of accelerating energetic charged particles in expanding 
relativistic outflows in a regime appropriate for Active Galactic Nuclei (AGN). It follows an earlier 
analysis where the implications for the high-speed (bulk flow Lorentz factors $\gamma_b \gppr 100$) 
gamma-ray-burst (GRB) regime has been investigated \cite{rieger05}, provides a general derivation 
of the relations presented there, and extends it to the AGN context.

In contrast to GRBs, AGN are in fact seen to exhibit relativistic outflows extending up to hundreds of
kilo-parsecs. In particular, for the "blazar" sub-class of AGN, radio VLBI/VLBA observations of their 
inner (pc-scale) jets frequently reveal significant apparent superluminal motion ($v_a>c$) of individual 
jet components propagating away from the core. When the fastest measured radio jet components are
put together, the velocity distribution shows a peak $\beta_a =v_a/c \sim 10$, with a tail extending up
to $\sim50$ \cite{Lister09,Piner12}, suggesting that (radio) flow speeds in AGN jets may reach bulk flow
Lorentz factors up to $\gamma_b \sim \beta_a/\beta \sim 50$. On the other hand, measured speeds of
the (parsec-scale) radio components in the VHE-detected, high-frequency-peaked BL Lac objects (HBLs) 
appear to be consistently well below those found in the above-noted, radio-selected samples \cite{Piner13}. 
If representative, the apparent lack of significant superluminal speeds in the parsec-scale radio 
jets of TeV-HBLs in fact would seem to suggest that the (radio) bulk flow Lorentz factor in these objects 
is only modest ($\gamma_b \sim 2-3$), in contrast to common constraints on their (sub-pc-scale) bulk 
Lorentz (Doppler) factors based on radiative modeling of their nuclear high-energy emission. In principle, 
such a difference could be accounted for by some change in directionality (e.g., jet bending, intrinsic helical 
motion) or by the presence of some velocity gradient in the flow, such as a (longitudinally) decelerating flow 
\cite{George03,Levinson07}, or a (transversally) structured jet with a fast-moving spine and a slower moving 
sheath \cite{Ghisellini05}. The former might be caused by radiative Compton drag, while the latter scenario 
of a fast inner flow (spine) encompassed by a slower outer flow (sheath) is generically expected in MHD 
models for the formation of relativistic jets with an ergospheric-driven jet surrounded by a slow moving disk 
wind \cite[see e.g.][]{McKinney06,Porth10,Hawley15}. The implied flow velocity gradients could possibly 
facilitate the acceleration of energetic charged particles to higher energies once, e.g., particles are efficiently 
scattered across the flow \cite{rieger06}, and this is what is studied here.

\section{Shear acceleration in spherical coordinates}
In the comoving frame $K'$, the acceleration coefficient in a gradual shear flow can be 
cast into the form \citep[e.g.,][eq.~3.27]{web89}
\beq\label{shear_coeff}
   \left<\frac{\Delta p\,'}{\Delta t'}\right>_{sh} = \frac{1}{p\,'^{\,2}}\frac{\pad }{\pad p\,'}
                \left(p\,'^{\,4}\,\tau'\,\Gamma\right)\,,
\eeq where $p\,'$ denotes the comoving particle momentum, with $p\,' \simeq p\,'\,^0$ for 
the energetic particles considered here, $\tau' \simeq \lambda'/c$ is the mean scattering 
time, and $\Gamma$ is the shear coefficient. In the strong scattering limit for quasi-isotropic
diffusion in a turbulent environment (i.e. $\omega_g\,'\,\tau' \lppr 1$, with $\omega_g\,'$ the 
relativistic gyro-frequency measured in the comoving frame) we have \citep[see][eq.~3.34]{web89}
\beq\label{viscous}
       \Gamma =\frac{c^2}{30}\,\sigma_{\alpha \beta}\,
       \sigma^{\alpha \beta}\,,
\eeq where $\sigma_{\alpha \beta}$, with $\alpha,\beta=0,1,2,3$, is the (covariant) fluid shear 
tensor given by \footnote{Note that this fixes a typographical sign error in eq.[A3] in \citep{rie04}.} 
\beq\label{shear}
       \sigma_{\,\alpha \beta}:=\nablaa u_{\beta}+\nablab u_{\alpha}
                 +\dot{u}_{\alpha} u_{\beta}+\dot{u}_{\beta} u_{\alpha}
                 -\frac{2}{3}\left(g_{\,\alpha \beta}+u_{\alpha}\,u_{\beta}
                  \right)\,\nabla_{\delta} u^{\delta}\,.
\eeq In this, $g_{\alpha \beta}$ denotes the (covariant) metric tensor and $\nabla_{\alpha}$ the 
covariant derivative. For spherical coordinates $x^{\alpha} = (c\,t,r,\theta,\phi)$, with $\phi$ 
the azimuthal and $\theta$ the polar angle, one has 
\beq
    (g_{\alpha \beta})=\rm{diag}(-1,1,r^2,r^2\,\sin^2\theta)\,.
\eeq
The only non-vanishing connection coefficients (Christoffel symbols of the second kind) are then 
given by
\beqn
   \Gamma^1_{22} & = & -r\,,\quad \Gamma^2_{21} =\Gamma^2_{12} 
                                   = \Gamma^3_{13} = \Gamma^3_{31} =\frac{1}{r}\,,
                            \quad \Gamma^2_{33}  =  -\sin\theta\,\cos\theta\,,\\
   \Gamma^1_{33} & = & -r\,\sin^2\theta\,,\quad \Gamma^3_{23} = \Gamma^3_{32}=\cot \theta\,.
\eeqn
Restricting ourselves to a time-independent, relativistic radial bulk flow velocity profile of the 
form
\beq
           u^{\alpha}= \gamma_b\,\left(1,v_r(r,\theta)/c,0,0\right)\,,
\eeq where $\gamma_b \equiv \gamma_b(r,\theta)=[1-v_r(r,\theta)^2/c^2]^{-1/2}$ is the bulk 
Lorentz factor, the fluid four divergence becomes
\beq\label{divergence}
  \nablab u^{\beta} = \frac{1}{r^2}\,\frac{\pad}{\pad r}\left(r^2 \gamma_b\,v_r/c\right)\,,
\eeq while the only non-vanishing components of the fluid four acceleration 
$\dot{u}_{\alpha} \equiv u^{\beta}\,\nablab\,u_{\alpha}$ are
\beqn
   \dot{u}_0 & = & -\gamma_b^4\,\frac{v_r^2}{c^3}\,\frac{\pad v_r}{\pad r}\,,\\
   \dot{u}_1 & = & \gamma_b^4\,\frac{v_r}{c^2}\,\frac{\pad v_r}{\pad r}\,.
\eeqn
For the non-vanishing components of the shear tensor one then finds
\beqn\label{shear-components1}
    \sigma_{00}&=&\frac{4}{3}\,\gamma_b^3\,\frac{v_r^2}{c^3}
                   \left(\gamma_b^2\,\frac{\pad v_r}{\pad r} - \frac{v_r}{r}\right)\,,\\
    \sigma_{01}&=&\sigma_{10} = - \frac{4}{3}\,\gamma_b^3\,\frac{v_r}{c^2}
                   \left(\gamma_b^2\,\frac{\pad v_r}{\pad r} - \frac{v_r}{r}\right)\,,\\
    \sigma_{11}&=& \frac{4}{3 c}\,\gamma_b^3 
                   \left(\gamma_b^2\,\frac{\pad v_r}{\pad r} - \frac{v_r}{r}\right)\,,\\
    \sigma_{12}&=&\sigma_{21} = \gamma_b^3\,\frac{1}{c}\,\frac{\pad v_r}{\pad \theta}\,,\\
    \sigma_{20}&=&\sigma_{02} = -\gamma_b^3\,\frac{v_r}{c^2}\,\frac{\pad v_r}{\pad \theta}\,,\\
    \sigma_{22}&=& \frac{2}{3}\,\gamma_b \,\frac{r^2}{c}
                   \left(\frac{v_r}{r} - \gamma_b^2\,\frac{\pad v_r}{\pad r}\right)\,,\\
    \sigma_{33}&=& \frac{2}{3}\,\gamma_b \,\frac{r^2}{c}\,\sin^2\theta
                   \left(\frac{v_r}{r} - \gamma_b^2\,\frac{\pad v_r}{\pad r}\right)\,.
\eeqn Noting that $\sigma^{\alpha \beta} = g^{\alpha \mu} g^{\delta \beta} \sigma_{\mu \delta}$,
the relativistic shear coefficient becomes
\beq\label{shear-coeff}
      \Gamma = \frac{4}{45}\,\gamma_b^2 \left[ \left(\gamma_b^2\,\frac{\pad v_r}{\pad r} - 
               \frac{v_r}{r}\right)^2 +\frac{3}{4\,r^2}\,\gamma_b^2\,\left(\frac{\pad v_r}{\pad \theta}
               \right)^2 \right]\,,
\eeq which for non-relativistic flow speeds (i.e., $\gamma_b \rightarrow 1$) and $v_r$ 
independent of $r$ (and $\phi$), i.e. $v_r \equiv v_r(\theta)$, reduces to 
\beq
      \Gamma = \frac{4}{45\,r^2} \left[v_r^2 + \frac{3}{4}\,\left(\frac{\pad v_r}{\pad \theta}
                                 \right)^2 \right]\,.
\eeq
It can be shown that this expression corresponds to the (non-relativistic) viscous transfer coefficient 
derived by Earl et al.~(1988) (their eq.~7) when the latter is expressed in spherical coordinates and 
the corresponding velocity profile $\vec{v} = v_r(\theta)\,\vec{e}_r$ is applied. 

For an energy-dependent scattering timescale of the form $\tau' \propto p\,'\,^{\alpha}$, the shear 
flow acceleration coefficient, eq.~(\ref{shear_coeff}), is given by
\beq\label{shear-coeff1}
   \left<\frac{\Delta p\,'}{\Delta t'}\right>_{sh} = (4+\alpha) \tau' \Gamma p'\,
\eeq so that the characteristic acceleration timescale $t_{\rm acc}(r,\theta) \simeq p\,'/<\dot{p}\,'>$ 
for gradual shear becomes
\beq\label{shear-timescale}
  t_{\rm acc}(r,\theta) \simeq \frac{45}{4 (4 + \alpha)}\frac{c}{\gamma_b^2\,\lambda'} 
                      \left[ \left(\gamma_b^2\,\frac{\pad v_r}{\pad r} - 
                      \frac{v_r}{r}\right)^2 
                      +\frac{3}{4\,r^2}\,\gamma_b^2\,
                      \left(\frac{\pad v_r}{\pad \theta}\right)^2\right]^{-1} ,
\eeq where in the presence of a background magnetic field the particle mean free path formally 
has to be smaller than the gyro-radius to satisfy the strong scattering.
Equation~(\ref{shear-timescale}) exemplifies the characteristic inverse dependence, $t_{\rm acc} 
\propto 1/\lambda'$, on the particle mean free path. This is related to the fact that in a shear flow
the average energy gain per scattering increases with increasing particle mean free path 
\cite[][]{rieger06}.

Consider the simplified case where the radial flow velocity is only a function of polar angle $\theta$, 
so that in four-vector notation the flow speed is given by 
\beq
   u^{\alpha}= \gamma_b\,\left(1,v_r(\theta)/c,0,0\right)\,,
\eeq where $\gamma_b \equiv \gamma_b(\theta)=[1-v_r(\theta)^2/c^2]^{-1/2}$ is the bulk 
Lorentz factor of the flow. The associated (comoving) timescale for the shear flow 
acceleration of particles then becomes 
\beq
    t_{\rm acc}(r,\theta) \simeq \frac{45}{4 (4 + \alpha)}\,
               \frac{c}{\lambda'}\,
                \frac{r^2}{\gamma_b^2\,\left[v_r^2 + 0.75\,\gamma_b^2\,   
                (\pad v_r/\pad \theta)^2\right]}\,,
\eeq where $r$ is the radial coordinate measured in the cosmological rest frame, $\lambda' 
\propto p'^{\alpha}$ is the particle mean free path, and $p'$ is the particle momentum in the 
comoving (jet) frame. As the jet flow is diverging and streamlines are separating, the acceleration 
timescale increases with the square of the radial coordinate $r$.\\

\section{Flow velocity profiles and related energy losses}
By means of application, let us consider three different bulk flow velocity profiles $v_r(\theta)=
\sqrt{1-1/\gamma_b(\theta)^2}$ parameterized in terms of $\gamma_b$ \cite[cf. also][for 
instantiation in the case of GRBs]{zha02,kum03,zha04}, i.e., a power-law model, where 
$\gamma_b$ is power-law function of $\theta$ outside a core of opening angle $\theta_c$, i.e., 
\beq
\gamma_b(\theta) = 1 + (\gamma_{b0} -1)\left(1+\left[\frac{\theta}{\theta_c}\right]^2\right)^{-b/2}\,,
\eeq with $1.5 < b \lppr 2$, a Gaussian profile with 
\beq
\gamma_b(\theta) = 1 + (\gamma_{b0} -1)~\exp\left({-\frac{\theta^2}{2\theta_c^2}}\right)\,,
\eeq and a Fermi-Dirac-type (top-hat) profile
\beq
\gamma_b(\theta) = 1 + (\gamma_{b0} -1) (1+\exp[-\beta_c]) /
              \left(1+\exp\left[\beta_c \left(\frac{\theta}{\theta_c}-1\right)\right]\right)
\eeq with $\beta_c > 0$, and where $\gamma_{b0}$ denotes the Lorentz factor at the jet 
axis ($\gamma_{b0} \lppr 50$ for AGN), and typically $\theta_c \lppr 0.1$ rad. Particle 
energization in these flow profiles then competes with conventional energy-loss processes.

\subsection{Adiabatic Losses}
\begin{figure}
\centering
\includegraphics[scale=.80]{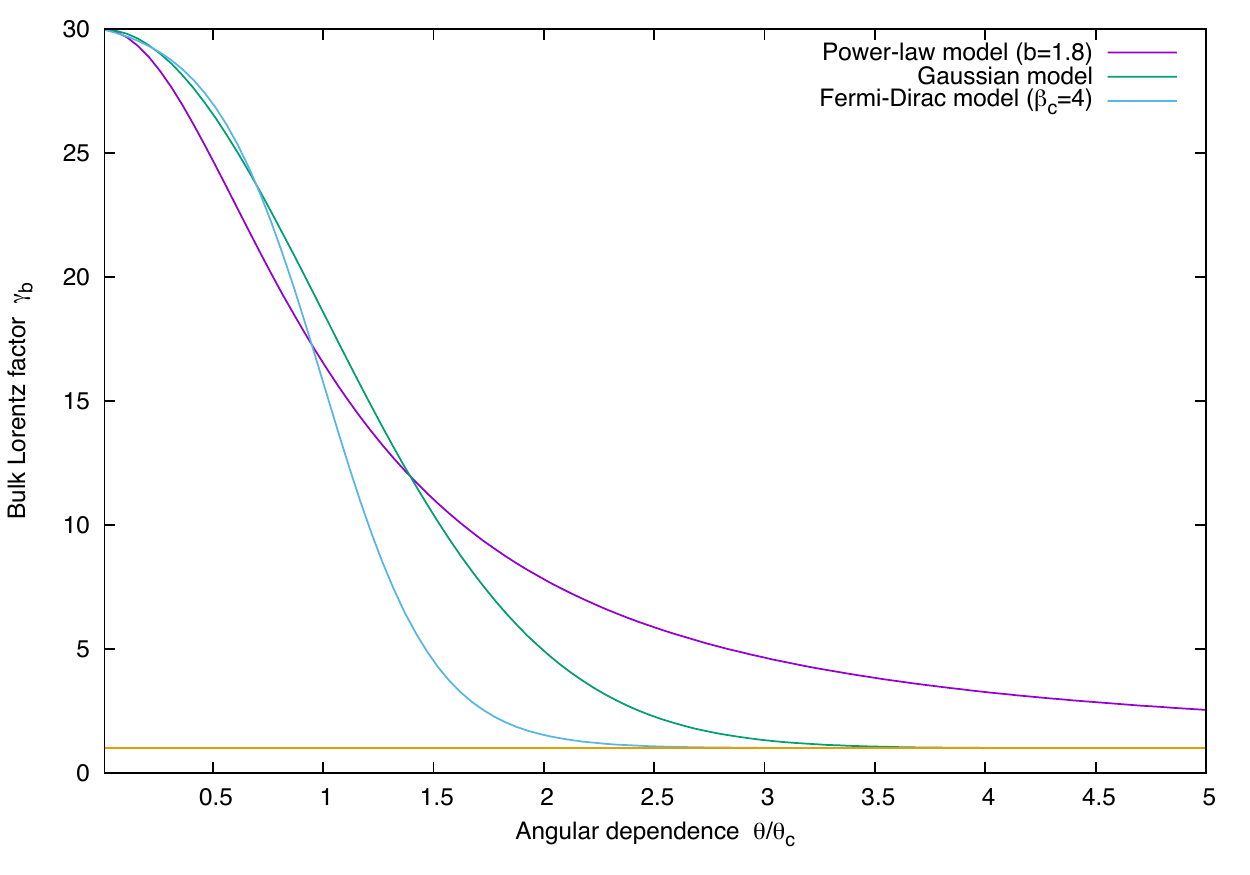}
\caption{Illustration of the evolution of the bulk Lorentz factor $\gamma_b$ with polar angle $\theta$ 
for a power-law ($b=1.8$), Gaussian and Fermi-Dirac type ($\beta_c=4$) profile, respectively, assuming 
$\gamma_{b0}=30$.\label{fig1}}
\end{figure}

\begin{figure}
\centering
\includegraphics[scale=.80]{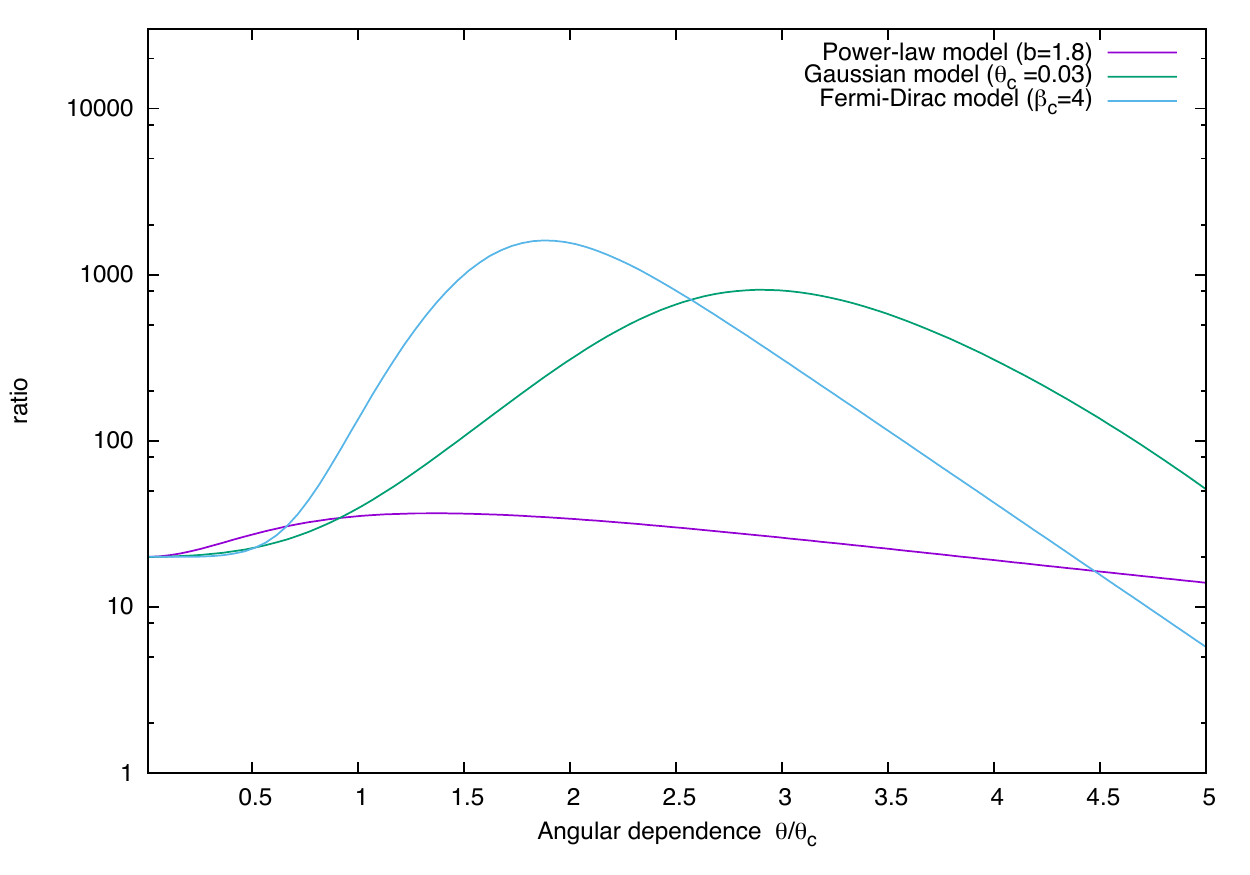}
\caption{Associated ratio $\frac{<\dot{p}'>_{sh}}{| <\dot{p}'>_{ad} |} \times \left(\frac{r}{\lambda'}\right)$ 
between energization by viscous shear and adiabatic losses for the profiles given in Fig.~\ref{fig1}, 
illustrated for $\theta_c=0.03$ [rad].\label{fig2}}
\end{figure}

For the corresponding adiabatic energy changes one finds (using eq.~\ref{divergence})
\beq
    \left<\frac{\Delta p\,'}{\Delta t'}\right>_{ad} := -\frac{p' c}{3} \nablab u^{\beta} 
        = - \frac{p'}{3} \left(2 \gamma_b \frac{v_r}{r} +\gamma_b^3 \frac{\partial v_r}{\partial r}\right) 
        = - \frac{2 p'}{3} \gamma_b \frac{v_r}{r}\, 
\eeq where the last relation holds for $v_r$ independent of $r$. This gives the ratio of viscous 
gain versus adiabatic losses to
\beq\label{adiabatic_ratio}
  \frac{\left<\frac{\Delta p\,'}{\Delta t'}\right>_{sh}}{| \left<\frac{\Delta p\,'}{\Delta t'}\right>_{ad} |}  
      = \frac{2}{15}\,(4+\alpha)\,\gamma_b \left(\frac{\lambda'}{r}\right) \left(\frac{v_r}{c} +
         \frac{3}{4}\,\gamma_b^2 \frac{c}{v_r}\left(\frac{1}{c}\frac{\partial v_r}{\partial \theta}\right)^2 \right)
\eeq Hence, one expects viscous shear energization in the diffusion regime ($\lambda' < r$) for the
present application (radially expanding flows, no azimuthal component) to be important only in the 
relativistic regime. Figures~\ref{fig1} and \ref{fig2} show examples of the flow profile and energization 
ratio assuming $\gamma_0=30$. For this case, at a given $r$, only particles with $\lambda' \gppr 
0.03~r$ (power-law profile), $\lambda' \gppr 0.002~r$ (Gaussian profile), and $\lambda' \gppr 10^{-3} r$ 
(Fermi-Dirac) can get efficiently accelerated. Efficient acceleration thus needs energetic seed particles 
and is usually difficult to achieve for electrons unless the magnetic field is weak.
%------------ FIGURE 1-------------
%-------------------------------------------
%------------ FIGURE 2---------------
%------------------------------------------

Depending on the shape of the velocity profile, particles are more easily accelerated (i.e., require
less injection energy) at different angular scales, i.e., not necessarily in the innermost ($\theta \lppr
\theta_c$) region. This becomes particularly apparent for the chosen Gaussian and the Fermi-Dirac 
type profile, where the shear gain to adiabatic loss ratio (Fig.~\ref{fig2}) peaks at $\theta \simeq
(2-3)~\theta_c$. This could in principle introduce more complex emission features (see below) and 
support, for example, some ridge-line structure or a limb-versus centrally-brightened morphology 
\cite[see e.g.][for recent exemplary findings in the context of 3C~84 and Cygnus A, 
respectively]{Nagai14,Bocc16} in cases where potential differences in Doppler boosting are 
effectively compensated by more efficient particle acceleration.\\ 
For illustration, Fig.~\ref{fig4} shows two possible (optically-thin) intensity maps ($I=\int \epsilon\, dl_{los}$, 
with  $dl_{los}$ the line of sight element) for a jet possessing a Fermi-Dirac bulk flow profile and inclined 
at viewing angle $i=10^{\circ}$ in the case where particle injection across the jet is kept constant or 
varied with polar angle approximately following Fig.~\ref{fig2}. For simplicity, it is assumed that the
number density of non-thermal electrons (with differential energy distribution described by a power 
law of index $\alpha=2$) is proportional to the plasma density $n_e \propto \rho \propto 1/z^2$ 
(where $z$ is the distance along the jet axis) and that the comoving emissivity is synchrotron type,
$\epsilon' \propto \rho' B'^{(\alpha+1)/2}$, with $\rho'=\rho/\gamma_b$, $B'=B/\gamma_b$ (perpendicular 
scaling, with $B(z)\propto 1/z$) and $\epsilon=D^{(\alpha+3)/2} \epsilon'$ and $D$ the Doppler factor. 
In such a case, efficient shear acceleration leads to the appearance of a more prominent off-axis (ridge-line) 
structure. 

If the above-mentioned conditions (Eq.~\ref{adiabatic_ratio}) are satisfied, acceleration always proceeds (up 
to a factor of the order of unity) on timescales shorter (faster) than the relevant dynamical timescale $t_d' \sim 
\frac{r }{c~\gamma_b}$. Hence, as a particle moves out along $r$, it will (in the absence of significant 
radiative losses) continue to get further accelerated to higher energies until it eventually leaves the 
flow by cross-field escape.

\subsection{Synchrotron Losses}
\begin{figure}[H]
\centering
%\plottwo{synchro_ratio}{compton_ratio}
\plottwo{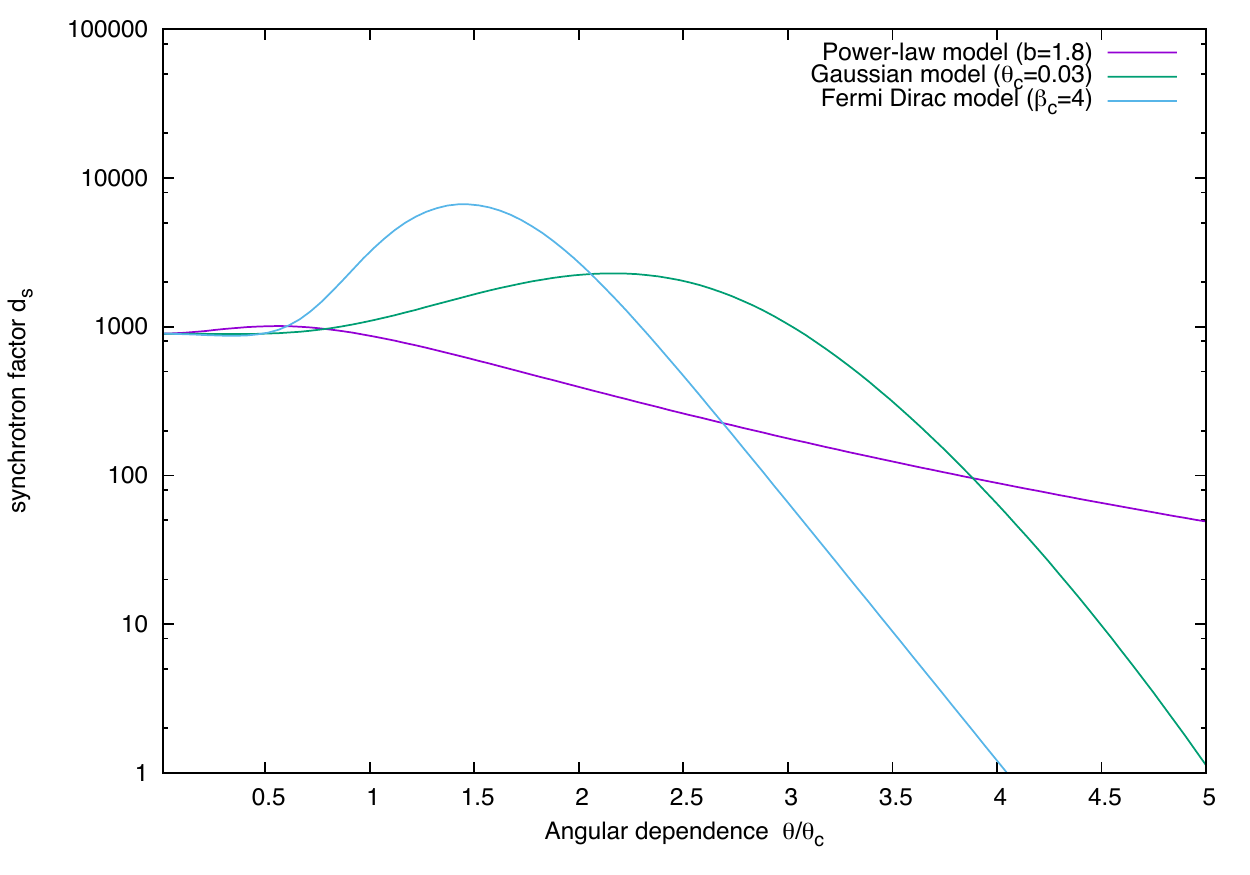}{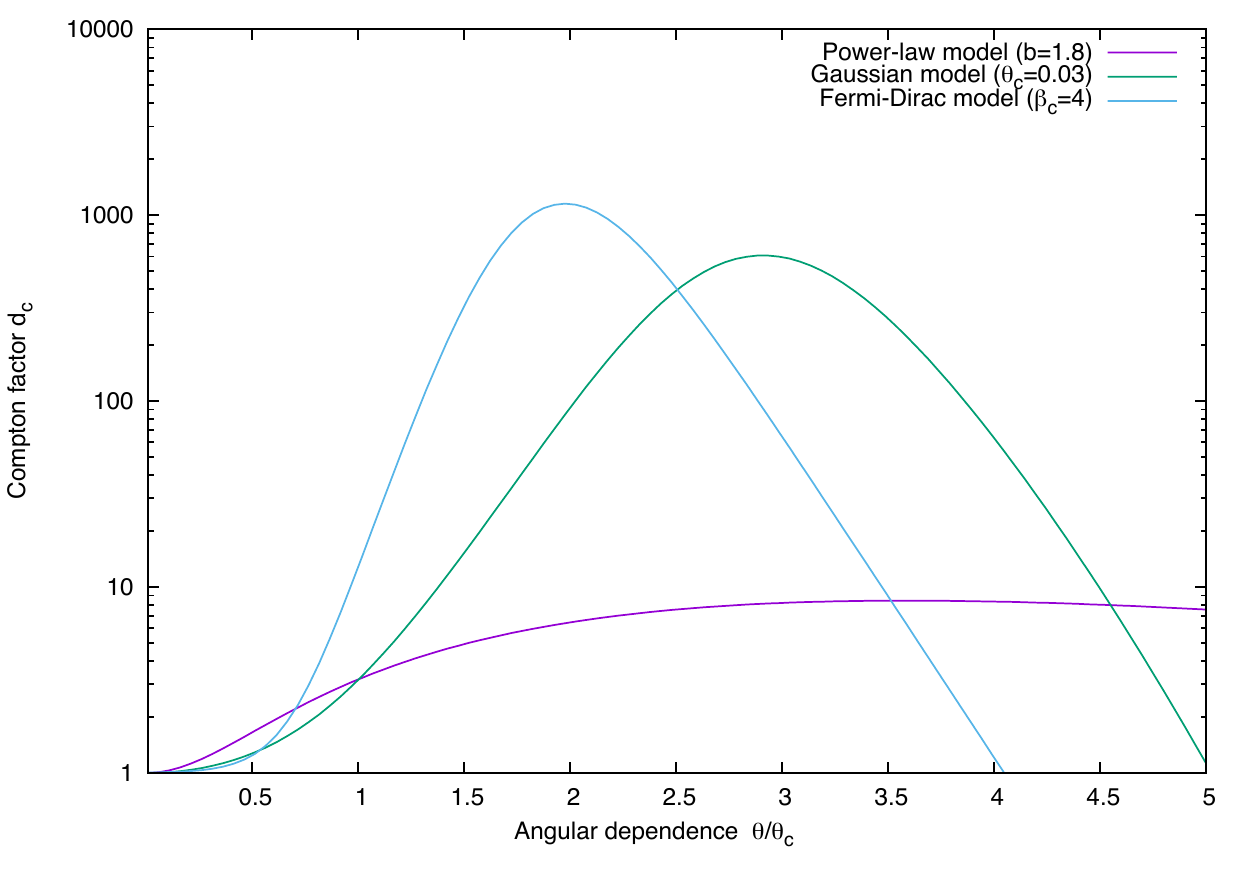}
\caption{Left: Associated synchrotron ratio factor $d_s$ for shear energization versus synchrotron 
cooling for the flow profiles given in Fig.~\ref{fig1}, illustrated for $\theta_c=0.03$ [rad].
Right: Associated Compton ratio factor $d_C$ for shear energization versus Compton cooling.
\label{fig3}}
\end{figure}

\begin{figure}[H]
\centering
%\plottwo{mathematica/intensity4_without}{mathematica/intensity4_with}
\plottwo{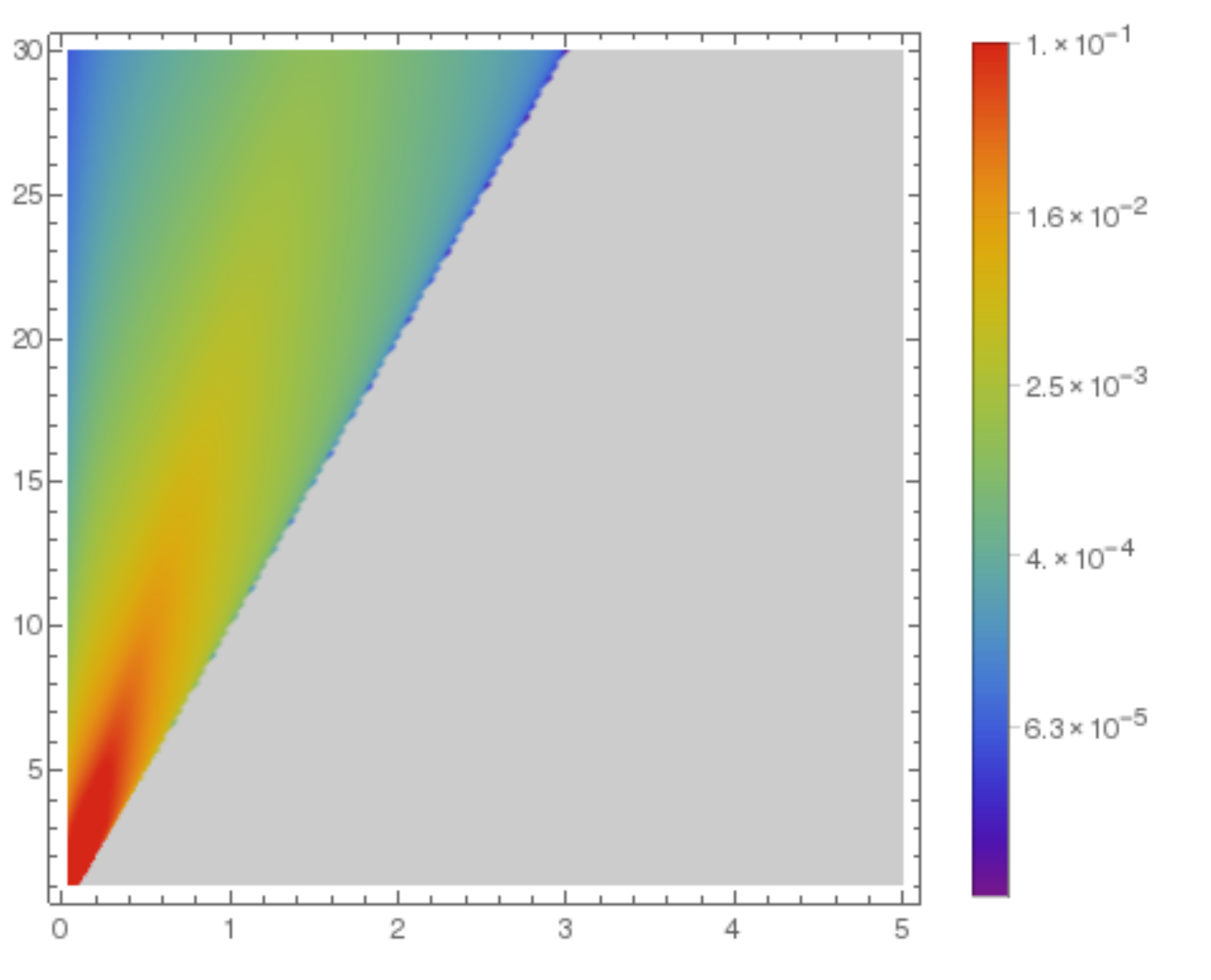}{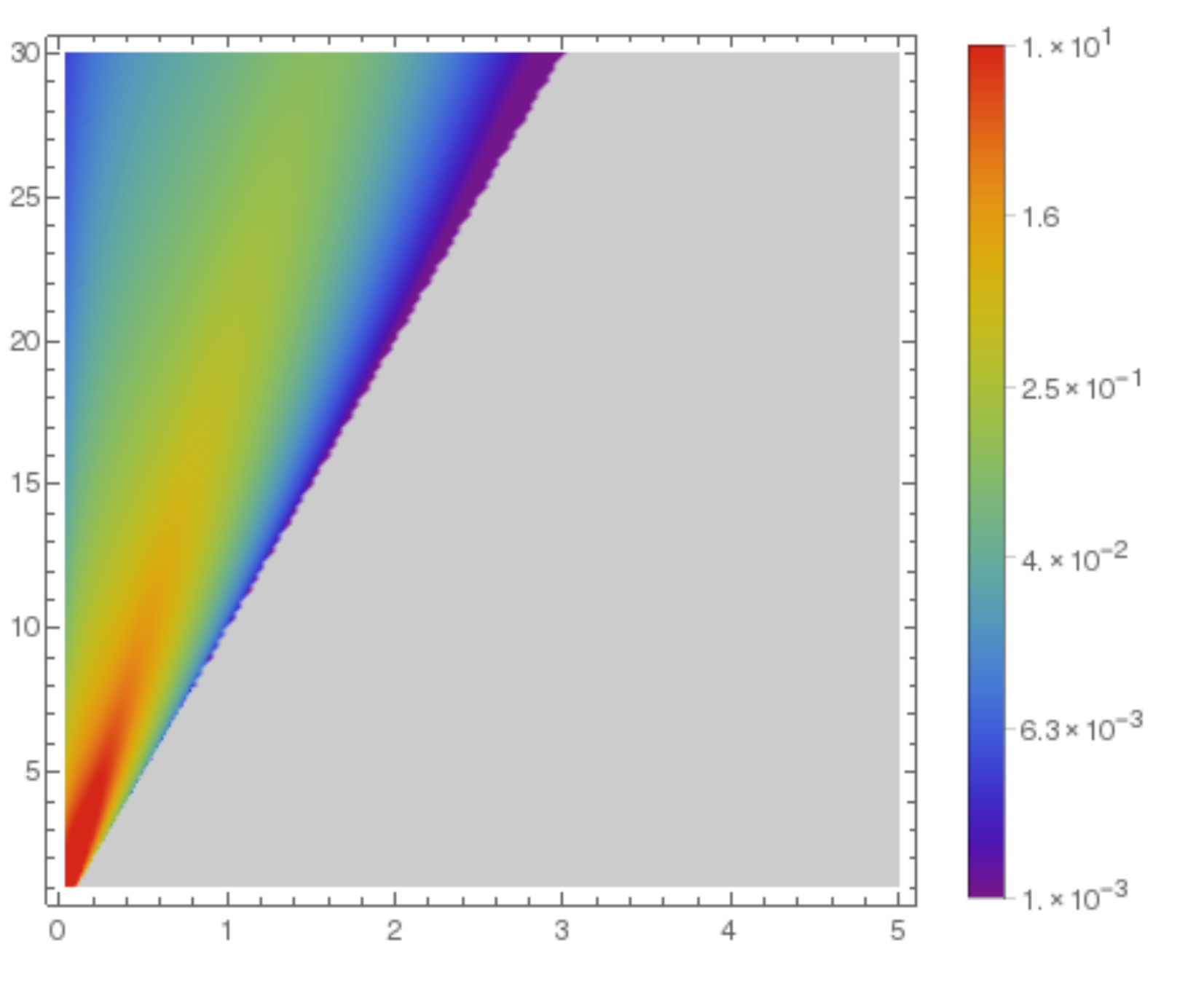}
\caption{Illustration of intensity maps showing the possible change in appearance for a relativistically 
expanding jet with a Fermi-Dirac type profile (using $\gamma_{b0}=30, \beta_c=4$ and $i=10^{\circ}$ 
for the inclination of the main $z$-axis to the line of sight) where the rate of particle injection is kept
constant ({\it left}) and varied ({\it right}) as a function (quasi-Gaussian centred around the peak value 
in Fig.~\ref{fig2}) of the polar angle $\theta$, respectively. 
\label{fig4}}
\end{figure}

In the presence of magnetic fields, energetic charged particles will undergo synchrotron losses
given by
\beq
  \left<\frac{\Delta p\,'}{\Delta t'}\right>_{\rm syn} = -\frac{4}{9} \frac{q^4}{m^2c^4}\,\gamma'^2 B'^2
\eeq with $m$ the particle's rest mass, $q=e$ its charge, and $\gamma'$ its Lorentz factor. We
assume the background magnetic field to scale with radius $B(r)' = B_0'~(r_0/r)^{a}$ with $0<a
\leq2$ (typically $a\sim 1$). The ratio of shear gain to synchrotron losses then becomes
\beq
  \frac{\left<\frac{\Delta p\,'}{\Delta t'}\right>_{sh}}{| \left<\frac{\Delta p\,'}{\Delta t'}\right>_{\rm syn} |} 
      = \frac{(4+\alpha)}{5} \left(\frac{\lambda'}{r_{gyro}'}\right) \frac{m^4 c^8}{e^5 B_0'^3} 
         \frac{1}{r_0^2} \gamma_b^2 \left(\frac{r}{r_0}\right)^{3a-2}
          \left(\frac{v_r^2}{c^2} +
           \frac{3}{4}\,\gamma_b^2 \left(\frac{1}{c}\frac{\partial v_r}{\partial \phi}\right)^2 \right)\,,
\eeq where $r_{gyro}' = \gamma' m c^2/(eB')$ is the comoving gyro-radius of the particle. Using 
characteristic (conical jet-type) scaling numbers in the AGN context (and $\alpha\simeq1$), this 
gives
\beqn
\frac{\left<\frac{\Delta p\,'}{\Delta t'}\right>_{sh}}{| \left<\frac{\Delta p\,'}{\Delta t'}\right>_{\rm syn} |} 
     & \simeq& 2 \times 10^{-10} \left(\frac{\lambda'}{r_{gyro}'}\right) \left(\frac{m}{m_e}\right)^4 
                  \left(\frac{10^3~\rm{G}}{B_0'}\right)^3
                  \left(\frac{10^{13}\rm{cm}}{r_0}\right)^2 \left(\frac{\gamma_b}{30}\right)^2 
                   \left(\frac{r}{r_0}\right)^{3a-2} \nonumber \\
     &\times&      \left(\frac{v_r^2}{c^2} +
                  \frac{3}{4}\,\gamma_b^2 \left(\frac{1}{c}\frac{\partial v_r}{\partial \theta}\right)^2 \right)\,.
\eeqn Hence, if $\lambda'$ scales with the gyro-radius, energetic protons ($m=m_p$) are expected 
to experience efficient acceleration right from the start, almost independently of the magnetic field 
scaling. On the other hand, for the chosen magnetic field dependence (only on $r$), electrons 
($m=m_e$) will only be efficiently accelerated if the magnetic field becomes sufficiently weak, 
e.g., for $a=1.5$ on scales $r \gppr 5\times 10^3 r_0$. This is illustrated in Fig.~\ref{fig3} where 
the synchrotron ratio factor $d_s:= \gamma_b^2 (v_r^2/c^2 + [3/4] [1/c^2] \gamma_b^2 
[\partial v_r/\partial \theta]^2)$ is plotted for the considered flow profiles. 
In reality, however, one may expect the magnetic field to also reveal some $\theta$-dependence, 
probably decreasing with $\theta$, thereby facilitating the acceleration of particles further away 
from the axis.
%------------ FIGURES 3 ---------------
%------------------------------------------

\subsection{Inverse Compton Losses}
For electrons, inverse Compton scattering (Thomson regime) could lead to additional cooling
effects,
\beq
  \left<\frac{\Delta p\,'}{\Delta t'}\right>_{\rm IC} = -\frac{4}{3} \sigma_T \gamma'^2 U_{ph}' \,,
\eeq where $\sigma_T$ is the Thomson cross-section and $U_{ph}'$ the energy density of the 
target photons as seen in the comoving frame. A non-reducible photon field is provided by the
CMB ($U_{CMB}=aT_{CMB}^4 \simeq 4.2 \times 10^{-13}$ erg cm$^{-3}$ $[1+z]^4$). With
$U_{ph}'=\gamma_b^2 U_{ph}$, the ratio of shear energization to Compton losses becomes
\beqn
  \frac{\left<\frac{\Delta p\,'}{\Delta t'}\right>_{sh}}{| \left<\frac{\Delta p\,'}{\Delta t'}\right>_{\rm IC} |} 
      &\simeq &  10^4 \left(\frac{\lambda'}{r_{gyro}'}\right) \left(\frac{r_0}{r}\right)^{2-a}
      \left(\frac{10^{13}\rm{cm}}{r_0}\right)^2 \left(\frac{10^3~\rm{G}}{B_0'}\right)
      \left(\frac{U_{CMB}}{U_{ph}}\right) \nonumber \\
       &\times& \left(\frac{v_r^2}{c^2} +
                     \frac{3}{4}\,\gamma_b^2 \left(\frac{1}{c}\frac{\partial v_r}{\partial \theta}\right)^2 \right)\,,
\eeqn using the same scaling for the magnetic field as above. The Compton ratio factor $d_C
:=(v_r^2/c^2 + [3/4] [1/c^2] \gamma_b^2 [\partial v_r/\partial \theta]^2)= d_s/\gamma_b^2$ is 
illustrated in Fig.~\ref{fig3}. Accordingly, once the magnetic field becomes sufficiently weak so 
that electron acceleration can overcome synchrotron losses ($<\dot{p}'>_{sh}/ | <\dot{p}'>_{syn} |
\propto r^{3a-2}$), shear energization is expected to continue out to larger distances until 
Compton losses set in ($<\dot{p}'>_{sh} / |<\dot{p}'>_{\rm IC} | \propto 1/r^{2-a}$).

\section{Discussion and Conclusion}
Shear flows are ubiquitous in astrophysical environments and potential sites of non-thermal
particle acceleration. Relativistic outflows are in fact known to be conducive to efficient Fermi
type shear particle acceleration \cite[e.g.,][]{Ostrowski2000,rie04}. In the present paper, we 
have explored the potential of expanding relativistic outflows to boost energetic particles to
higher energies. To  this end, a set of simplified (azimuthally symmetric) conically expanding 
flow profiles (power-law, Gaussian and Fermi-Dirac-type) has been examined where the 
outflow bulk Lorentz factor is solely a function of polar angle. When applied to the AGN context, 
the results show that for the acceleration mechanism to overcome radiative and non-radiative
losses and to work efficiently, the injection of pre-accelerated seed particles is required. This 
could in general be achieved by first-order shock or stochastic second-order Fermi processes. 
In this sense, shear acceleration would resemble a two-stage process for further particle 
energisation beyond the common limit. Depending on the shape of the flow profile, particles 
are more easily accelerated (i.e., require less injection energy) at different angular scales, i.e., 
not necessarily in the innermost core region close to the axis. This could in principle introduce 
different jet emission features (e.g., core versus off-axis ridge-line structures or limb-brightening) 
and allow for a variety in jet appearance. Once operative, gradual shear acceleration proceeds 
on a timescale inversely proportional to the particle mean free path, $t_{\rm acc} \propto 1/\lambda'$. 
For a gyro-dependent particle mean free path, $\lambda' \propto \gamma'$, this gives the same 
scaling as synchrotron losses, $t_{\rm acc}/t_{\rm syn} = {\rm const.}$ so that once started, 
synchrotron losses will not be able to further constrain particle acceleration, while for electrons 
Compton losses in an expanding flow might perhaps do so on larger scale.\\
From a methodological point of view, numerical studies focusing on the diffusive transport and 
acceleration of particles in turbulent fields \citep[e.g.][]{osullivan09} and the excitation of a 
large-scale shear dynamo \citep[e.g.][]{yousef08} become of particular interest to adequately 
quantify the potential of shear particle acceleration on the relevant scales.\\
We note that shear acceleration in AGN jets seems in principle capable of accounting for continued 
acceleration and related extended emission. The inverse dependence on the particle mean free 
path makes shear acceleration a preferred mechanism for the acceleration of hadrons and provides 
some further weight to the relevance of AGN jets for our understanding of the origin of the recently 
detected high-energy (PeV) neutrinos \cite[e.g.][]{Becker14,Tavecchio15} and the production of 
extreme cosmic-rays \cite[e.g.][]{Lemoine13}.

%variability = dynamical timescale

%------------ FIGURES 4 ---------------
%------------------------------------------

\acknowledgements
F.M.R. kindly acknowledges support by a DFG Heisenberg Fellowship (RI 1187/4-1).

\clearpage


\begin{thebibliography}{}
\bibitem[Aloy \& Mimica 2008]{aloy08} Aloy, M.A., and Mimica, P. 2008, ApJ, 681, 84
\bibitem[Becker Tjus et al. 2014]{Becker14} Becker Tjus, J., Eichmann, B., Halzen, F. et al. 2014, PhRvD, 89, 123005 
\bibitem[Bell 2013]{Bell13} Bell, A.R. 2013, APh, 43, 56
\bibitem[Boccardi et al. 2016]{Bocc16} Boccardi, B. et al. 2016, A\&A, 585, 33 % Cygnus A jet stratification !
\bibitem[Earl et al. 1988]{ear88} Earl, J.A., Jokipii, J.R., \& Morfill, G.\ 1988, \apj, 331, L91 
\bibitem[Georganopoulos \& Kazanas 2003]{George03} Georganopoulos, M. and Kazanas 2003, D., ApJ, 594, L27  % Decelerating flows in TeV blazars  !
\bibitem[Ghisellini et al. 2005]{Ghisellini05} Ghisellini, G., Tavecchio, F. and Chiaberge, M. 2005, A\&A, 432, 401 % Radiative, spine-shear interplay in TeV BL Lac  !
\bibitem[Grismayer et al. 2013]{grismayer13} Grismayer, T., Alves, E.P., Fonseca, R.A. and Silva, L.O. 2013, PPFC, 55, 124031 
\bibitem[Hawley et al. 2015]{Hawley15} Hawley, J.F., Fendt, C., Hardcastle, M. et al. 2015, SSRv, 191, 441 % Jet disk connection review !
\bibitem[Jokipii \& Morfill 1990]{jok90} Jokipii, J.R., \& Morfill, G.E. 1990, \apj, 356, 255
\bibitem[Kumar \& Granot 2003]{kum03} Kumar, P., \& Granot, J. 2003, \apj, 591, 1075 
\bibitem[Laing \& Bridle 2013]{laing13} Laing, R.A. and Bridle, A.H. 2013, MNRAS, 432, 1114 % Jet base spectra in FR I : implications for particle acceleration !
\bibitem[Lemoine 2013]{Lemoine13} Lemoine, M. 2013, JPhCs, 409, 2007 % Acceleration and propagation of ultrahigh energy cosmic rays !
\bibitem[Levinson 2007]{Levinson07} Levinson, A. 2007, ApJ, 671, L29 % Radiative deceleration In TeV blazars.  !
\bibitem[Liang et al. 2013]{liang13} Liang, E., Boettcher, M. and Smith, I. 2013, ApJ, 766, L19 % m.f. generation and shear acceleration !
\bibitem[Lister et al. 2009]{Lister09} Lister, M.L. et al. 2009, AJ, 138, 1874 %MOJAVE blazar jet - superluminal motion  !
\bibitem[McKinney 2006]{McKinney06} McKinney, J.C. 2006, MNRAS, 368, 1561 % MHD jets from rapidly rotating BHs  !
\bibitem[Nagai et al. 2014]{Nagai14} Nagai, H. et al. 2014, ApJ, 785, 53 % limb-brightening in 3C84 !
\bibitem[Ohira 2013]{ohira13} Ohira, Y. 2013, ApJ, 767, 16 
\bibitem[Ostrowski 2000]{Ostrowski2000} Ostrowski, M. 2000, MNRAS, 312, 579 % On possible `cosmic ray cocoons' of relativistic jets !
\bibitem[Piner et al. 2012]{Piner12} Piner, B.G. et al. 2012, ApJ, 758, 84 % superluminal motion & acceleration in blazar jets  !
\bibitem[Piner \& Edwards 2013]{Piner13} Piner, B.G., and Edwards, P.G. 2013, EPJWC 6104021 % slow motion in newer TeV blazars  !
\bibitem[Porth \& Fendt 2010]{Porth10} Porth, O. and Fendt, C. 2010, ApJ, 709, 1100  % MHD disk winds  !
\bibitem[Rieger \& Duffy 2004]{rie04} Rieger, F.M., \& Duffy, P. 2004, \apj, 617, 155
\bibitem[Rieger \& Duffy 2005]{rieger05} Rieger, F.M., \& Duffy, P. 2005, \apj, 632, L21
\bibitem[Rieger \& Duffy 2006]{rieger06} Rieger, F.M., \& Duffy, P. 2006, \apj, 652, 1044
\bibitem[Rieger \& Aharonian 2012]{rieger12} Rieger, F.M. and Aharonian, F.A. 2012, MPLA, 27, 12300301 
\bibitem[Sahayanathan 2009]{Sahayanathan09} Sahayanathan, S. 2009, MNRAS, 398, 49 % Shear acceleration in Mkn 501 !
\bibitem[O'Sullivan et al. 2009]{osullivan09} O'Sullivan, S., Reville, B., \& Taylor, A.~M.\ 2009, \mnras, 400, 248 
\bibitem[Tavecchio \& Ghisellini 2015]{Tavecchio15} Tavecchio, F., \& Ghisellini, G.\ 2015, MNRAS, 451, 1502 % High-energy cosmic neutrinos from spine-sheath BL Lac jets
\bibitem[Webb 1989]{web89} Webb, G.M. 1989, \apj, 340, 1112
\bibitem[Yousef et al. 2008]{yousef08} Yousef, T.~A., Heinemann, T., Rincon, F., et al.\ 2008, Astronomische Nachrichten, 329, 737 % numerical large-scale, shear dynamo studies 
\bibitem[Zhang \& M{\' e}sz{\' a}ros 2002]{zha02} Zhang, B., \& M{\' e}sz{\' a}ros, P.\ 2002, \apj, 571, 876 
\bibitem[Zhang et al. 2004]{zha04} Zhang, B., Dai, X., Lloyd-Ronning, N.M., \& M{\' e}sz{\' a}ros, P.\ 2004, \apj, 601, L119 
\end{thebibliography}
\end{document}